\documentclass[a4paper]{jpconf}
\usepackage{iopams}
\usepackage{graphicx}
\newcommand{\mi}{\mathrm{i}}

\begin{document}
\title{Vortices in exciton-polariton condensates with polarization splitting}
\author{M Toledo Solano and Yuri G Rubo}
\address{Centro de Investigaci\'on en Energ\'{\i}a,
 Universidad Nacional Aut\'onoma de M\'exico, \\
 Temixco, Morelos, 62580, Mexico}

\ead{ygr@cie.unam.mx}

\begin{abstract}
The presence of polarization splitting of exciton-polariton branches in
planar semiconductor microcavities has a pronounced effect on vortices
in polariton condensates. We show that the TE-TM splitting leads to the
coupling between the left and right half-vortices (vortices in the
right and left circular components of the condensate), that otherwise
do not interact. We analyze also the effect of linear polarization
pinning resulted from a fixed splitting between two perpendicular
linear polarizations. In this case, half-vortices acquire strings
(solitons) attached to them. The half-vortices with strings can be
detected by observing the interference fringes of light emitted from
the cavity in two circular polarizations. The string affects the
fringes in both polarizations. Namely, the half-vortex is characterized
by an asymmetric fork-like dislocation in one circular polarization;
the fringes in the other circular polarization are continuous, but they
are shifted by crossing the string.
\end{abstract}

\section{Introduction}
Half-integer vortices (HV) are elementary topological defects in
superfluids and Bose-Einstein condensates with multicomponent order
parameter \cite{VolovikBook}. They are characterized by half-quantum
change of the phase of the condensate, i.e., the phase is changed by
$\pm\pi$ after encircling the point of singularity. In condensates of
exciton-polaritons in planar semiconductor microcavities, where the
order parameter has two components, namely, the left and the right
circular-polarization components, the vortex, in general, is described
by two winding numbers $(k,m)$, where $k$ is for the polarization
angle, and $m$ is for the phase. The HV is characterized by
simultaneous change of both angles by $\pm\pi$, so that $k,m=\pm 1/2$
and there are four different HV's \cite{Rubo2007}. Higher-order vortex
entities can be considered as superpositions of several HV. For
example, the integer phase vortex $(0,1)$ is the superposition of two
HV, $(1/2,1/2)$ and $(-1/2,1/2)$. Both the integer phase vortices
\cite{Lagoudakis2008} and the half-integer vortices
\cite{Lagoudakis2009} have been discovered recently in
exciton-polariton condensates.

The experimental observations \cite{Lagoudakis2008,Lagoudakis2009}
revealed the importance of polarization splitting of polariton branches
and, in particular, the presence of pinning of the condensate
polarization to a specific crystallographic direction. The HV can be
observed only when this splitting is sufficiently small. The goal of
this paper is to present the brief theoretical discussion of the
effects of weak polarization splitting on the properties of HV's. In
what follows we discuss two different types of splitting, the
wave-vector dependent TE-TM splitting and the wave-vector independent
splitting due to the cavity asymmetry.

\section{Vortices in presence of TE-TM splitting}
Near the bottom of lower polariton branch the kinetic energy density of
polariton condensate is
\begin{equation}
 \label{KinEnergy}
 \mathfrak{T}=\frac{\hbar^2}{2}\sum_{i,j=x,y}
 \left\{
 \frac{1}{m_t}
 (\nabla^{\vphantom{*}}_i\psi^*_j)(\nabla^{\vphantom{*}}_i\psi^{\vphantom{*}}_j)
 +\left(\frac{1}{m_l}-\frac{1}{m_t}\right)
 (\nabla^{\vphantom{*}}_i\psi_i^*)(\nabla^{\vphantom{*}}_j\psi^{\vphantom{*}}_j)
 \right\},
\end{equation}

 \vspace{-0.1\baselineskip}\noindent
where $m_t$ and $m_l$ are the transverse (TE) and longitudinal (TM)
effective masses of polaritons. The two-dimensional vector
$\bpsi=\{\psi_x,\psi_y\}$ is the condensate wave function normalized to
the condensate concentration $n=|\bpsi|^2$. The presence of TE-TM
splitting does not destroy HV's, but it results in the warping of their
polarization texture. More importantly, it changes qualitatively the HV
energies and HV interactions. We consider these effects below for the
case of uniform exciton-polariton condensate without any potential
disorder.

In the elastic region, where the condensate concentration is constant,
the order parameter of linearly polarized condensate can be written as
$\bpsi=\sqrt{n}\,e^{\mi\theta}\{\cos\eta,\sin\eta\}$, where $\theta$
and $\eta$ are the phase and polarization angles, respectively. In the
case of small difference $|m_l-m_t|$, one can take these angles as in
unperturbed case. Namely, in the polar coordinates $(r,\phi)$ these
angles depends on the azimuth as $\theta=m\phi$ and $\eta=k\phi$. The
kinetic energy density then reads
\begin{equation}
 \label{KinEnergy2}
 \mathfrak{T}=\frac{\hbar^2n}{2m^*}\frac{(k^2+m^2)}{r^2}
 +\frac{\hbar^2n}{4}\left(\frac{1}{m_l}-\frac{1}{m_t}\right)
 \frac{(k^2-m^2)}{r^2}\cos[2(k-1)\phi],
\end{equation}
\begin{equation}
 \label{MStar}
 \frac{1}{m^*}=\frac{1}{2}\left(\frac{1}{m_l}+\frac{1}{m_t}\right).
\end{equation}

 \vspace{-0.1\baselineskip}
The energy of single vortex in logarithmic approximation is found by
integrating this expression over the microcavity plane. The integral
should be cut at small $r$ by the vortex core radius $a$, and for large
$r$ by the excitation spot radius $R$. The core radius is
$a=\hbar/\sqrt{2m^*\mu}$, where the chemical potential $\mu$ can be
measured by the blue-shift of the emission line of the polariton
condensate. The value of $a$ is typically about a few microns. Due to
the oscillating factor, the second term in (\ref{KinEnergy2})
contributes only for $k=1$, and the energy of the vortex is
\begin{equation}
 \label{SingleEng}
 E_{(k,m)}=
 \frac{\pi\hbar^2n}{2}\left[
  \left(\frac{1}{m_l}+\frac{1}{m_t}\right)(k^2+m^2)
 +\left(\frac{1}{m_l}-\frac{1}{m_t}\right)(1-m^2)\delta_{1,k}
                     \right]\ln\left(\frac{R}{a}\right).
\end{equation}

 \vspace{-0.2\baselineskip}
It is seen, that the energies of most of the vortices are defined by
$m^*$, but the energy of polarization ``hedgehog'' vortex $(1,0)$ is
defined be the pure longitudinal mass $m_l$. This result can be
visualized by consideration of polarization texture of integer vortices
shown in figure~1. For the $(0,\pm1)$ and $(-1,0)$ vortices
[figure~1(a-c)] the polarization field is longitudinal in some areas
and transverse in others. For the $(1,0)$ vortex [figure~1(d)] the
field is longitudinal everywhere.

Without TE-TM splitting there is no long-range interaction between the
left and right HV's, that is between HV's with the different sign of
the product $km$ \cite{Rubo2007}. The TE-TM splitting leads to weak
coupling for a particular pair, the $(1/2,1/2)$ and $(1/2,-1/2)$
half-vortices. This can be easily understood since when these HV's are
put together they form the ``hedgehog'' vortex with the energy
$E_{(1,0)}=(\pi\hbar^2n/m_l)\ln(R/a)$. When this pair are well
separated the energy of the system is equal to the doubled energy of
the single HV, that is to $(\pi\hbar^2n/m^*)\ln(R/a)$. The interaction
energy of these vortices separated by distance $r_{12}$ is
\begin{equation}
 \label{InterEng}
 V_{12}=\frac{\pi\hbar^2n}{2}\left(\frac{1}{m_t}-\frac{1}{m_l}\right)
   \ln\left(\frac{r_{12}}{a}\right).
\end{equation}

 \vspace{-0.2\baselineskip}
The long-range interaction between left and right HV's can be important
for the analysis of the Berezinskii-Kosterlitz-Thouless (BKT)
transitions (without this interaction there are two decoupled BKT
transitions \cite{Rubo2007,Keeling2008}). It should be noted also that
the interrelation between $m_t$ and $m_l$ depends on the detuning of
the frequency of the cavity photon mode from the center of the
stop-band of the distributed Bragg mirror \cite{Panzarini1999}. So, one
can have both weak attraction and weak repulsion of the $(1/2,1/2)$ and
$(1/2,-1/2)$ half-vortices.

\begin{figure}[t]
\includegraphics[width=0.5\textwidth]{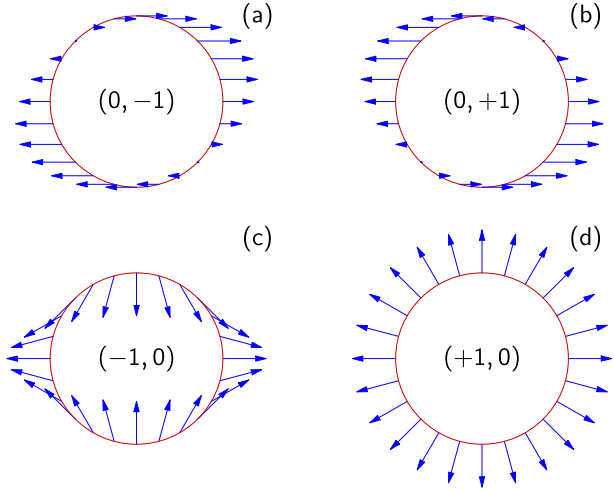}\hspace{0.05\textwidth}%
\begin{minipage}[b]{0.45\textwidth}
 \caption{\label{Fig-IntVorts}
 Showing the polarization texture of four different integer vortices.
 Arrows indicate the instant directions of the order parameter. The value of
 the phase is reflected in the arrow length. The field is mixed
 longitudinal-transverse for the vortices (a-c), but it is purely longitudinal
 for the ``hedgehog'' vortex (d).}
\end{minipage}
\end{figure}

\section{Half-vortices with strings and interference fringes}
The TE-TM splitting is absent for position-independent order parameter.
In this section we consider the effects of the other type of
polarization splitting observed in exciton-polariton condensates. In
this case the energy of a uniform condensate becomes dependent on the
orientation of polarization. The corresponding term in the condensate
energy density is \cite{Kasprzak2007}
\vspace{-0.25\baselineskip}
\begin{equation}
 \label{SplitEnergy}
 \mathfrak{H}_\epsilon=\frac{\epsilon}{2}\left(n-|\psi_x|^2+|\psi_y|^2\right)
 =\epsilon n\sin^2(\eta),
\end{equation}
where $\epsilon$ is the energy of the splitting and we have chosen the
$x$-axis along the direction of polarization pinning (the easy axis for
polarization), so that the energy of the condensate is minimized for
the horizontal polarization. Below we consider the case of weak
pinning, when the energy of splitting $\epsilon\ll\mu$ and the
characteristic length $b=\hbar/\sqrt{2m^*\epsilon}\gg a$.

In conditions of \emph{cw} excitation one can expect that the HV will
be formed with the polarization texture that minimizes the total
elastic energy. The minimization of
$\int(\mathfrak{T}+\mathfrak{H}_\epsilon)d^2r$ for the case $m_t=m_l$
gives the equations for the phase and polarization angles,
\vspace{-0.25\baselineskip}
\begin{equation}
 \label{SineGordon}
 \Delta\theta=0, \qquad 2b^2\Delta\eta=\sin(2\eta).
\end{equation}

\vspace{-0.25\baselineskip}
The solution to the first equation
describes the same uniform change of the phase angle with the azimuth,
$\theta=m\phi$ with $m=\pm1/2$. However, the polarization angle $\eta$
changes from 0 to $\pm\pi$ only in a narrow region $\sim b$ when one
encircles the core far away from the center. This region of rapid
change of $\eta$ defines a string or soliton attached to the HV (see
figure 16.1 in \cite{VolovikBook}). Formally, the string is defined by
the line where polarization becomes vertical.

The numerical solution to the sine-Gordon equation (\ref{SineGordon})
is shown in figure~2(a). It was calculated with the boundary condition
$d\eta/d\phi=k$ near the core ($r\ll b$) with the winding number
$k=1/2$. The axes are chosen such that $x$-axis is directed to the
right and $y$-axis is directed downwards, so that the azimuth $\phi$
increases for the clockwise rotation. For large $r\gg b$, the gradient
of $\eta$ is zero almost everywhere, except the region of large
negative $y$ and $|x|\sim b$, where the angle changes according to the
known kink solution $\eta=2\tan^{-1}[\exp(x/b)]$. This way the solution
in figure~2(a) describes the HV with the string going upwards.

In general, any orientation of the string is possible and HV's with the
strings oriented differently have the same energy for the case
$m_t=m_l$. Note that the string carries the energy proportional to its
length. In finite size condensates the string can terminate on the
boundary. In this case the HV will be attracted by the boundary and can
relax towards it. The string can terminate also in another HV with the
opposite sign of the winding number $k$, and this leads to the
interaction of left and right HV's. The interaction energy between HV's
with opposite signs of $k$ coupled by the string becomes proportional
to the length of the string and grows linearly with the distance.

The behavior of the interference fringes in two circular polarizations
is shown in figure~2(b,c) for the string terminating into the
$(1/2,1/2)$ half-vortex. These fringes appear when one observes the
interference of the signal emitted from the polariton condensate in a
circular polarization with plane wave of the same intensity propagating
in the downward direction. Experimentally
\cite{Lagoudakis2008,Lagoudakis2009}, this plane wave originates from
the same condensate but from a different place, where the order
parameter is approximately constant. The signal emitted in
$\sigma^{\pm}$ polarization is $\propto\exp\{\mi(\theta\mp\eta)\}$,
respectively, and the fringes in figure~2(b,c) were calculated as
$|e^{\mi(\theta\mp\eta)}+e^{\mi\kappa y}|^2$ with $\kappa=2/b$.

One can see that the half-vortex with the string is characterized by an
asymmetric fork-like dislocation in one circular polarization.
Moreover, the fringes that cross the string are shifted by half
wave-length in both circular polarizations. The choice of orientation
of the plane wave-vector to be collinear with the string (so that the
fringes are perpendicular to it) is the best to observe the string in
the interference pattern.

\vspace{-0.25\baselineskip}
\ack We are grateful to Alexey Kavokin,
Jonathan Keeling, and Igor Luk'yanchuk for discussions. This work was
supported in part by the grant IN107007 of DGAPA-UNAM.

\begin{figure}[t]
\begin{center}
\includegraphics[width=\textwidth]{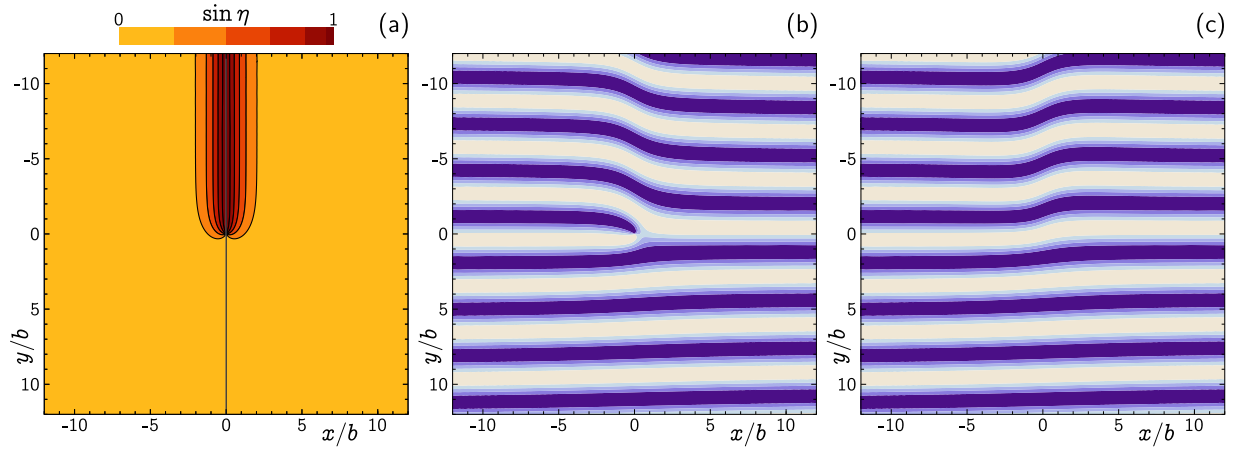}
\end{center}
 \caption{\label{Fig-EtaFringes}
 (a) The contour plot showing the change of the polarization angle $\eta$ for HV
 with the string going vertically up. The solid lines are drawn for the values of
 $\eta$ changing from 0 to $\pi$ (on the line going downwards) with the step $\pi/12$.
 Parts (b) and (c) show the interference fringes (see text) in the left-circular ($\sigma^-$)
 and in the right-circular ($\sigma^+$) polarizations, respectively.}
\end{figure}

\vspace{-0.25\baselineskip}

\end{document}